\documentclass[showpacs,prb,twocolumn]{revtex4}

\usepackage{graphicx}
\usepackage{amsmath}
\usepackage{color} 
\usepackage{transparent}
\usepackage{hyperref}

\def\bra#1{\mathinner{\langle{#1}|}}
\def\ket#1{\mathinner{|{#1}\rangle}}

\def\bog {B$_{1g}$ }
\def\aog {A$_{1g}$ }

\begin{document}
\title{The effect of electron correlation on the electronic structure and spin-lattice coupling of the high-T$_c$ cuprates: quantum Monte Carlo calculations}
\author{Lucas K. Wagner}
\email{lkwagner@illinois.edu}
\author{Peter Abbamonte }
\affiliation{Dept. of Physics, University of Illinois at Urbana-Champaign}
\begin{abstract}
Electron correlation effects are particularly strong in the high temperature superconducting materials. 
Devising an accurate description of these materials has long been a challenge, with these strong correlation effects historically being considered impossible or impractical to simulate computationally. 
Using quantum Monte Carlo techniques, we have explicitly simulated electron correlations in several cuprate materials from first principles.
These simulations accurately reproduce many important physical quantities about these materials, including the interaction-induced gap and the superexchange coupling between copper spins, with no additional parameters beyond fundamental constants.
We further investigate the dimensionless spin-lattice coupling parameter in the parent materials, showing that it varies dramatically between 0.1 and 1.0 depending on the interlayer.
This result indicates that the lattice and magnetic degrees of freedom are not independent in these systems, which may have ramifications for the origin of superconductivity.
\end{abstract}

\maketitle


The phenomenon of high temperature superconductivity in the copper oxides has been a decades long challenge to fully describe.
The phase diagram is very complicated, with structural transitions, magnetic transitions, and metal-insulator transitions occuring in close proximity to one another.
There are indications of strong coupling of electrons with some other degree of freedom\cite{gweon_unusual_2004} that may be magnetic or structural in origin, and there is a puzzling isotope effect at low doping\cite{pringle_effect_2000,crawford_anomalous_1990} that disappears around optimal doping.
These indications of electron-phonon coupling\cite{gunnarsson_interplay_2008} are seemingly in contradiction to the strong evidence for a magnetic origin for superconductivity summarized recently in Scalapino~\cite{scalapino_common_2012}. 
It has been proposed that spin and lattice can act cooperatively to enhance superconductivity\cite{yin_correlation-enhanced_2013}, which provides a compelling impetus to completely understand the magneto-structural coupling in the cuprates.
It is thus clear that spin, charge, and lattice degrees of freedom are active in the phase space near the superconducting state, but their precise roles are still controversial.

Fully exploring the interactions between spin and lattice has stymied both experimental and theoretical efforts.
Experimentally, most techniques can only probe one of these degrees of freedom directly, and the magnetic excitations and lattice degrees of freedom are near to one another energy, making it challenging to disentangle their effects on each other, although there has been notable progress in that area\cite{conte_disentangling_2012}.
From the theoretical standpoint, accurately describing the electronic structure of the cuprates has been a tremendous challenge because of the strong effects of electron correlation present in these materials.
Standard density functionals fail on a qualitative level\cite{pickett_fermi_1992}, particularly in the insulating undoped system, requiring {\it a posteriori} corrections\cite{anisimov_band_1991} that rob the method of predictive power. It has been shown that the electronic structure calculated in standard density functional theory(DFT) is not reliable even for electron-phonon interactions\cite{reznik_photoemission_2008,pintschovius_electron-phonon_2005}.

\begin{figure}
  \includegraphics[width=\columnwidth]{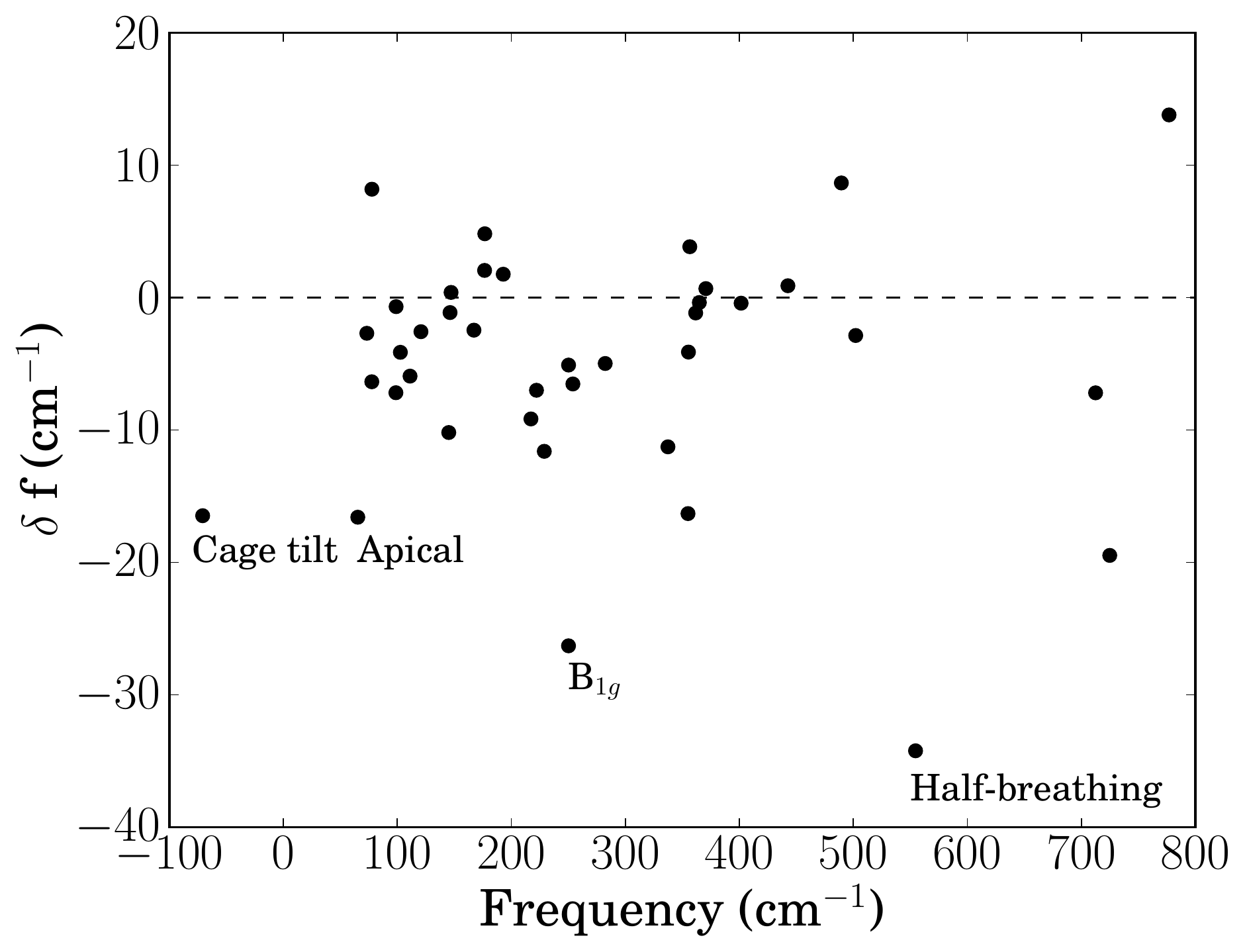}
  \caption{The change in frequency upon switching from a checkerboard antiferromagnetic ordering to ferromagnetic ordering for all $\Gamma$ phonons for La$_2$CuO$_4$ calculated with the PBE0 density functional. The modes with large changes in frequency are labeled: a) `Cage tilt' is the rotation of the oxygen cage towards the orthorhombic structure b) `Apical' is a mode involving the apical oxygen atoms 
    c) `\bog' is the d-wave oxygen buckling mode, and 
  d) `Half-breathing' is the out of phase breathing mode of the oxygen atoms in the CuO$_2$ plane.}
  \label{fig:phonon_spectrum}
\end{figure}

In this article, we elucidate the spin-lattice coupling in the cuprates by performing first-principles quantum Monte Carlo (QMC) calculations of several instances of the CuO$_2$ plane present in the cuprates: the real materials La$_2$CuO$_4$ and CaCuO$_4$, and a hypothetical unsupported CuO$_2^{2-}$ plane.
Unlike density functional methods, the QMC calculations explicitly treat electron correlation within these strongly correlated materials, which allows us to make a clear assessment of the importance of this physics to the basic electronic structure.
We examine the effect of explicit correlations on the spin-lattice coupling and focus on understanding the difference between the s-wave \aog and d-wave \bog oxygen buckling modes.
The \bog symmetry oxygen buckling mode has been studied closely in models\cite{gunnarsson_interplay_2008} and experiment\cite{reznik_q_1995,friedl_determination_1990,pyka_superconductivity-induced_1993,baron_first_2008,bakr_electronic_2009,raichle_highly_2011,cuk_review_2005}.
Experimentally, the \bog mode shifts and broadens on entering the superconducting state, while the closely related \aog-symmetry oxygen buckling mode does not.
Also, the \bog mode has also been implicated in dispersion kinks in ARPES.\cite{cuk_coupling_2004}
However, the closely related \aog mode does not seem to be strongly affected. 
A straightforward hybrid DFT calculation (Fig~\ref{fig:phonon_spectrum}) shows that the \bog mode has a very large reaction to a change in the magnetic state, which could be an explanation for the shifts in mode; however, as we mentioned before, it is not clear that the DFT calculations are reliable in these materials.
We find that in contrast to DFT calculations, state of the art QMC methods can accurately predict many properties of the cuprates, including the gap, magnetic coupling, and phonon frequencies.
The \aog and \bog modes are differentiated through interaction with the interlayer.

We apply the approximate, but highly accurate fixed-node diffusion Monte Carlo (FN-DMC) as implemented in the QWalk package\cite{wagner_qwalk:_2009}.  
We start with one-particle orbitals from a density functional calculation from CRYSTAL2009\cite{dovesi_crystal:_2005}, using pseudopotentials based on Dirac-Fock methods\cite{lee_pseudopotentials_2000,burkatzki_energy-consistent_2007,dolg_combination_1993}.
A Slater determinant is then constructed from the one-particle orbitals and multiplied by a Jastrow correlation factor, which is variance optimized. 
Finally, the FN-DMC method is performed with the resulting Slater-Jastrow wave function as a guiding function using t-moves\cite{casula_beyond_2006}.
The general procedure is outlined in the literature~\cite{foulkes_quantum_2001,wagner_energetics_2007,wagner_transition_2007} and enough details to reproduce the results, including basis sets and pseudopentials, are presented in the supplementary information.
We also provide a table of the raw energies in the supplementary information, for reference.

\begin{figure*}
  \begin{tabular}{ccc}
    \includegraphics[width=0.6\columnwidth]{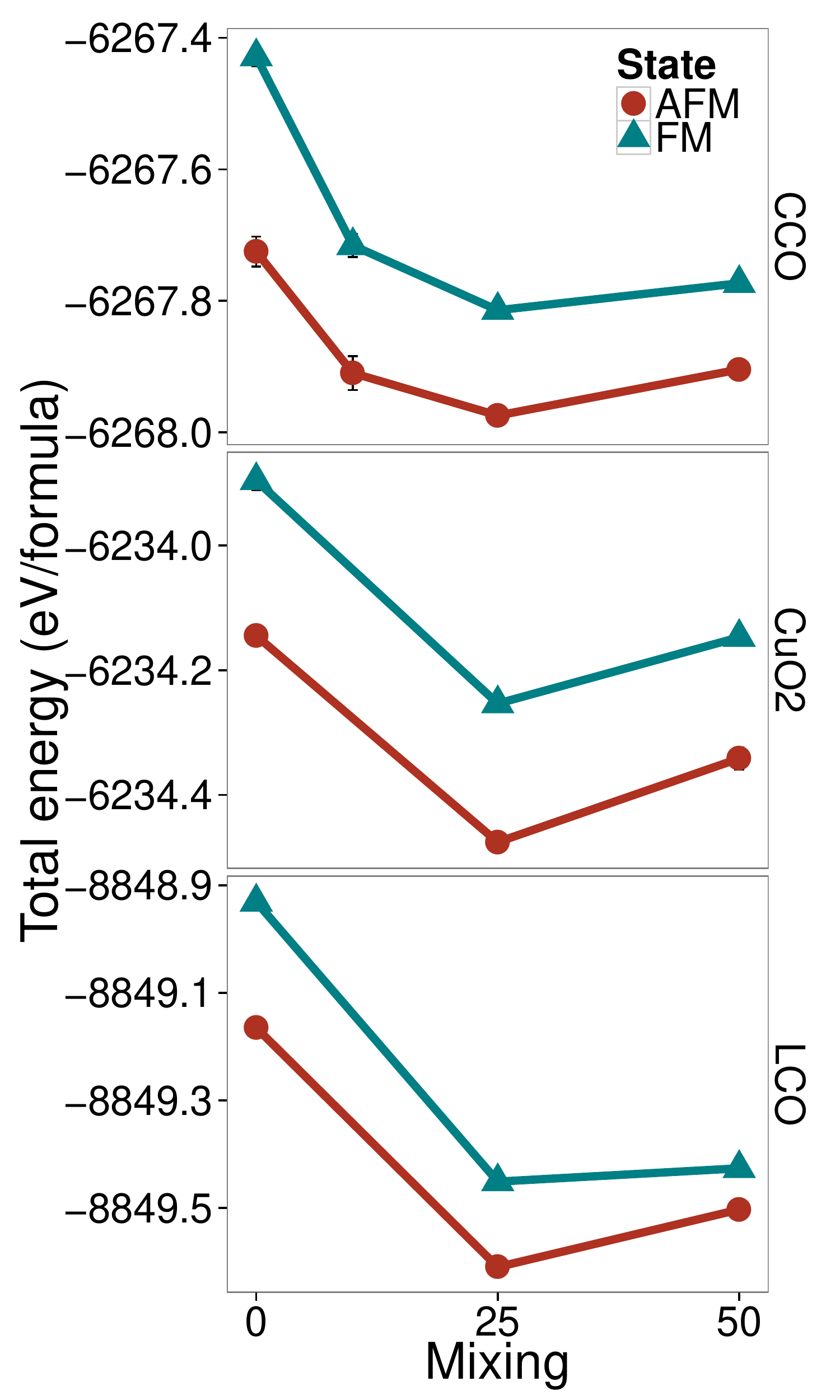} & 
    \includegraphics[width=0.6\columnwidth]{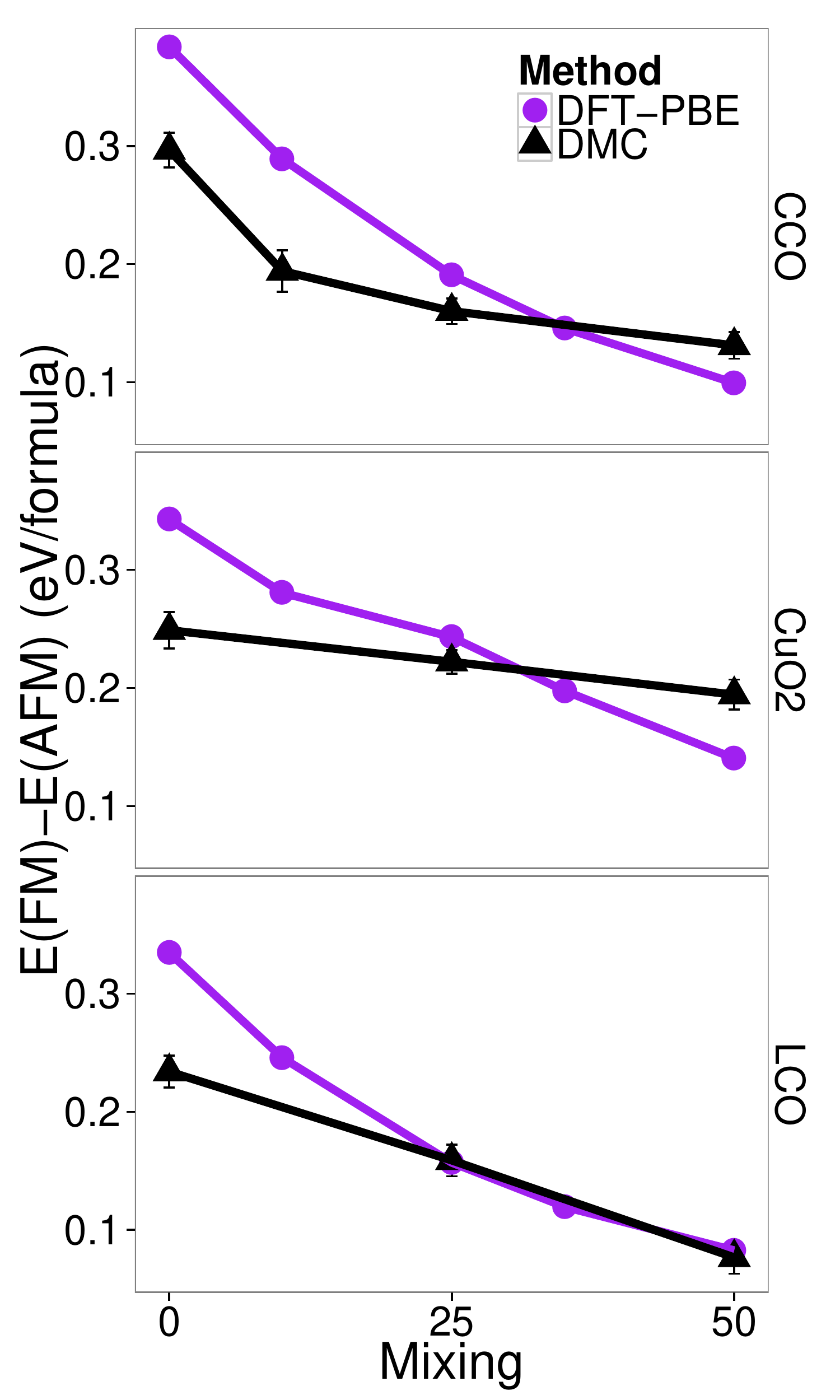} &
  \includegraphics[width=0.6\columnwidth]{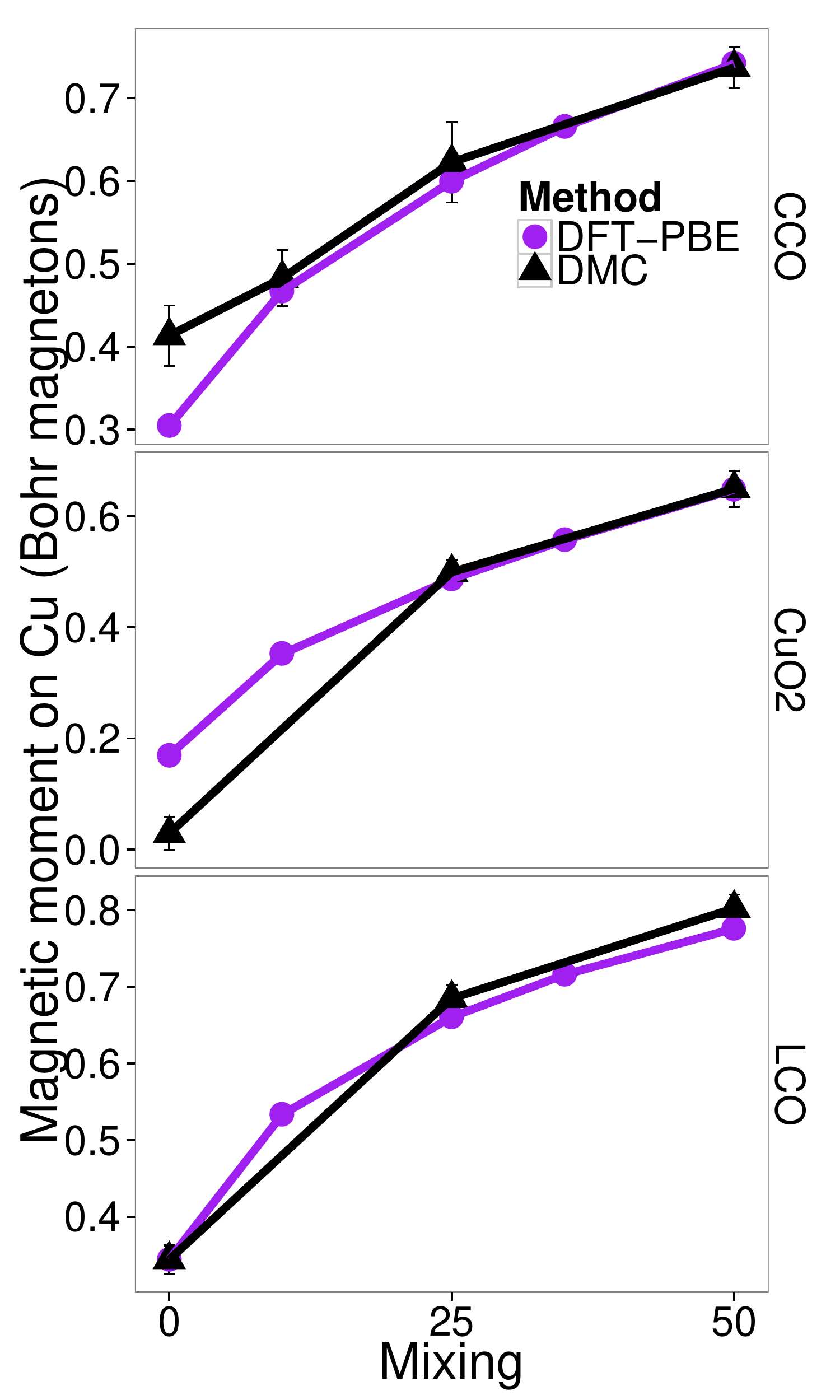} \\
  {\bf (a)} & {\bf (b)} & {\bf (c)} \\
\end{tabular}
\caption{(color online) Controlling the fixed node error by varying the one-particle orbitals.
  {\bf (a) } Total FN-DMC energy as a function of the mixing in density functional theory used to generate the orbitals.
  {\bf (b) } Energy difference between different magnetic states as a function of the orbitals.
  {\bf (c) } Estimated magnetic moments as a function of the orbitals.
  Lines are guides to the eye, and error bars, when larger than the symbol size, represent one standard deviation of the statistical uncertainty.
}
  \label{fig:allvsmixing}
\end{figure*}

There are four major approximations in the FN-DMC method: the systematic time step and finite size error, and the methodological fixed node and pseudopotential error.
We used a time step of 0.02 Hartree$^{-1}$; we checked that reducing the time step to 0.01 Hartree$^{-1}$ did not change the results within error bars.
We used highly accurate Dirac-Fock pseudopotentials designed for many-body calculations with explicit 3s and 3p electrons to reduce the effect of the pseudopotential approximation.
Finally, we tested larger supercell sizes than $2\times2\times1$ by expanding both to the $2\times2\times2$ and $2\sqrt{2}\times2\sqrt{2}\times1$ cell sizes for the CaCuO$_2$ case, and found no changes to our estimated $J$ within error bars.

The major approximation in the FN-DMC method is the nodal error; we check the dependence of our results on the Slater determinant used in the trial function.
It has been observed previously\cite{kolorenc_wave_2010,wagner_energetics_2007} that for transition metal oxides, the fixed node error can be improved by using a hybrid DFT functional to generate the orbitals.
The main effect of changing the exact exchange mixing is to change the relative position of the $d$ and $p$ orbitals, which then affects the hybridization.
Since FN-DMC is variational in the total energy, the most accurate wave function is the one with the lowest energy.
In Fig~\ref{fig:allvsmixing}a, we present the total energy as a function of the exact exchange mixing used to generate the orbitals. 
The minimum is near 25\% for all materials considered here, which is similar to other transition metal oxides.

In Figs~\ref{fig:allvsmixing}b and \ref{fig:allvsmixing}c, we present the dependence of the energy difference between the ferromagnetic and antiferromagnetic states and the magnetic moment on the nodal surface.
The results from the corresponding density functional theory are shown for comparison.
There are two main points to take away from these tests. 
First is that the fixed node condition does not allow the magnetic moment to change very much from the trial wave function.
The second is that despite this, the energy differences are corrected significantly from the DFT, and generally are much less sensitive to the input trial wave function.
We variationally select the nodal surface with the lowest energy, which is obtained by using an exact exchange mixing of 25\%.
This is equivalent to optimizing the trial Slater determinant in a reduced space.
Fully optimizing the Slater determinant would be ideal, but is currently computationally infeasible.

\begin{table}
\caption{The \bog and \aog modes refer to the oxygen buckling modes in the tetragonal symmetry labeling.
The hypothetical \aog mode used for comparison here is not an eigenmode of the dynamical matrix, and thus does not have an experimental value.}
\begin{tabular}{l|c|c|c|c}
   Quantity & PBE\footnote{For consistency, we estimated the PBE J by forcing AFM and FM magnetic orders and taking the energy difference. PBE predicts an unpolarized state at lower energy than the FM state, so the implicit Heisenberg model does not technically apply in that case. PBE0 and FN-DMC do not suffer from this abiguity.}  & PBE0 & FN-DMC & Experiment \\
\hline
\hline
\multicolumn{5}{c}{La$_2$CuO$_4$}  \\
\hline
J (eV)                       & 0.34&0.16& 0.160(13) & 0.11\cite{coldea_spin_2001} \\
$\mu$ (Bohr)                 & 0.36&0.69& 0.68(2)  & 0.6 \\
Optical gap (eV)             & 0.33&3.86& 2.8(3) & 2.2 \\
AFM B$_{1g}$ freq (meV)      & 30.1  & 30.3 & 32(1)   & 33\cite{wang_first-principles_1999} \\
FM B$_{1g}$ freq (meV)       & 24.6  & 26.3 & 25(2)   &\\
AFM A$_{1g}$ freq (meV)      & 42.8  & 45.1 & 45(1) & \\
FM A$_{1g}$ freq (meV)       & 39.9  & 43.7 & 43(1)   &  \\
\hline
\multicolumn{5}{c}{CaCuO$_2$}  \\
\hline
J (eV)                      & 0.13 & 0.18 & 0.14(2) & .14\cite{kan_preparation_2004}  \\
$\mu$ (Bohr)     & 0.33 & 0.60 & 0.61(2) & 0.61(2)\cite{kan_preparation_2004} \\
Optical gap (eV)            & 0.00 & 2.9 & 2.4(2)\footnote{Corrected by 0.6 eV because we evaluated the $\Gamma\rightarrow\Gamma$ transition}  & 1.8\cite{kan_preparation_2004} \\
AFM B$_{1g}$ freq (meV)     & -5.7 & -6.5 & 13(1) & $>0$\\
FM B$_{1g}$ freq (meV)      & -10.5 & -17.8 & -6(3) &\\
AFM A$_{1g}$ freq (meV)     & 21.7 & 24.5 & 26(2) & \\
FM A$_{1g}$ freq (meV)      & 13.4 & 20.6 & 24(2) & \\
\hline
\multicolumn{5}{c}{CuO$_2^{(2-)}$}  \\
\hline
J (eV)                   & 0.05 & 0.25 & 0.24(1) & \\
$\mu$ (Bohr)  & 0.17 & 0.50 & 0.50(2) &  \\
AFM B$_{1g}$ freq (meV)  & -22.7 & -22.1 & -20(2)  &  \\
FM B$_{1g}$ freq (meV)   & -29.9 & -29.0 &  -27(1) & \\
AFM A$_{1g}$ freq (meV)  & 16.5 & 20.1   & 25(2) & \\
FM A$_{1g}$ freq (meV)   & 5.9 & 11.4   & 17(2) &  \\
\end{tabular}
\label{table:la2cuo4}
\end{table}

The results of the FN-DMC procedure in comparison to experiment and two common density functionals are summarized in Table~\ref{table:la2cuo4}.
As one can see from the comparison table, the hybrid functional PBE0 overestimates the gap significantly, while FN-DMC is able to describe the gap and the exchange coupling accurately without any fitting.
As has been noted before\cite{giustino_small_2008}, the DFT-calculated frequency of the phonon modes is in fairly good agreement with experiment, with the exception of the \bog mode in CaCuO$_2$. 
Both DFT functionals obtain a negative frequency for this mode, which is inconsistent with experimental structures.
Our FN-DMC results are also able to reproduce the phonon frequencies quite accurately.
It is important to emphasize that these results were obtained {\em without} adjustable parameters and explicit simulation of the electron interactions.
To our knowledge, with the recent exception of the calculation of J\cite{foyevtsova_ab_2014} using similar techniques, this has not been achieved for so many physical quantities for the cuprates.

\begin{figure}
  \includegraphics[width=\columnwidth]{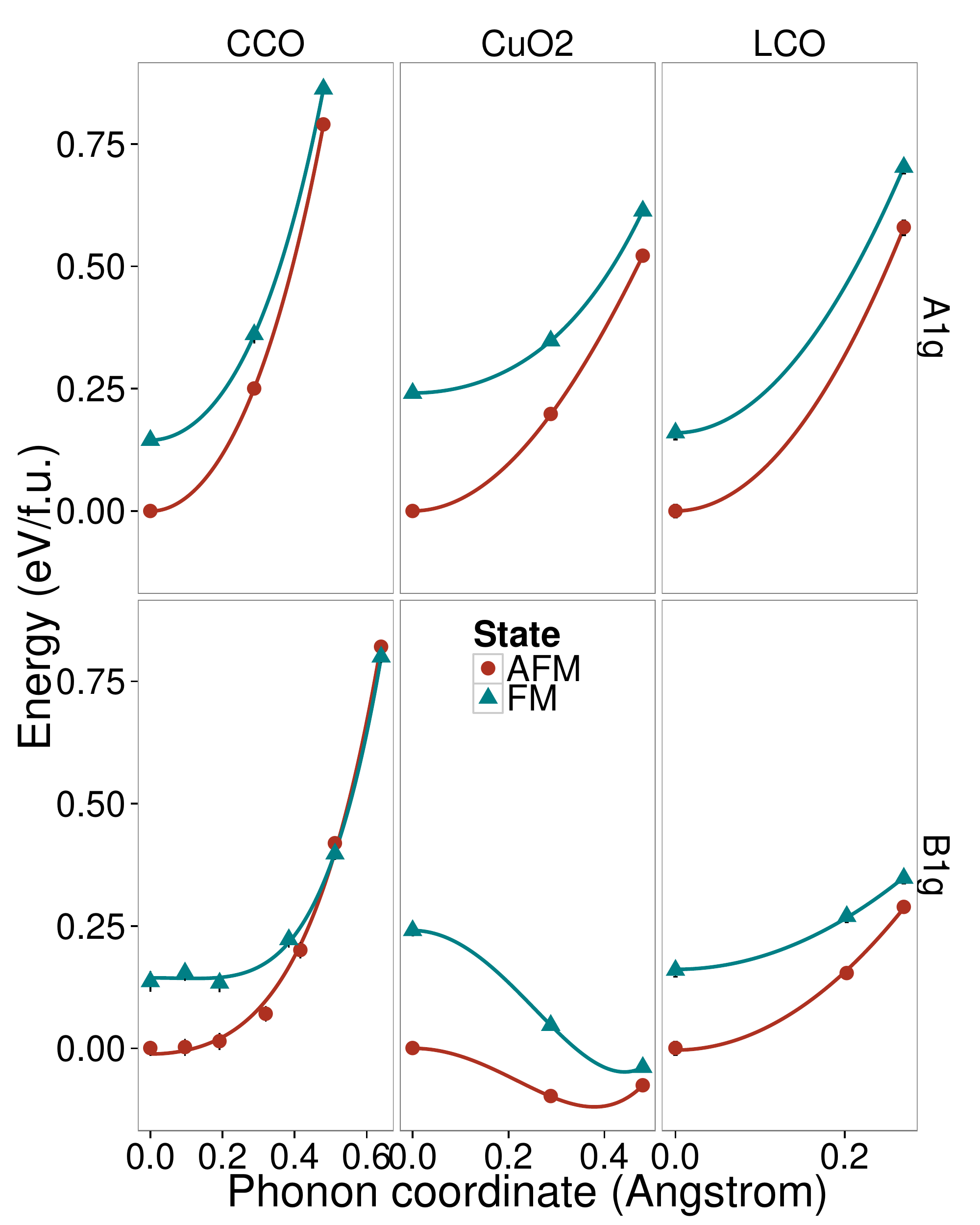}
  \caption{The energy as a function of the phonon coordinate for all systems considered in this work.}
  \label{fig:energy_as_u}
\end{figure}

We calculate the Heisenberg superexchange parameter $J$ by comparing the electronic state constrained to have all copper moments spin-aligned (FM) to the spin-anti-aligned moments in the checkerboard pattern (AFM).
It is worth noting that we are fitting to a classical Heisenberg model in this case.
This is because our simulations have collinear spins, and thus do not include the non-collinear fluctuations in the quantum Heisenberg model.
To check the validity of the Heisenberg model, we also calculated the energy of the mixed moment alignment (stripe) state, with aligned moments along the $a$ direction and anti-aligned moments along the $b$ direction.
We found that $E_{FM}(u)-E_{AFM}(u)=2(E_{\text{stripe}}-E_{AFM})$ in FN-DMC within error bars and in PBE0, justifying the use of the Heisenberg model.
Two oxygen buckling modes are considered: \aog where all oxygen atoms move in phase, and \bog, which has a d-wave alternating pattern of oxygen displacements.
As mentioned above, the \bog mode has been implicated in a number of interesting phenomena in the cuprates.
To assess the importance of the interlayer, we consider three different systems: La$_2$CuO$_4$ with apical oxygen atoms, CaCuO$_2$, which lacks apical oxygen atoms, and the isolated copper-oxide plane CuO$_2^{2-}$.

For a given frozen phonon distortion $u$, we calculate the superexchange parameter as a function of the phonon coordinate $u$:
\begin{equation}
  J(u)=J_0 + \delta c_1 u^2 + \delta c_2 u^4 \propto E_{FM}(u)-E_{AFM}(u).
\end{equation}
For La$_2$CuO$_4$, we set $c_2=0$ and checked that within DFT the quadratic approximation was a good fit.
For CaCuO$_2$ and CuO$_2^{2-}$, the quartic model was fit.
The energy calculations and the fits are shown in Fig~\ref{fig:energy_as_u}.


Given $J(u)\simeq J_0+\delta c_1u^2+\delta c_2u^4$, we fit to spin-lattice coupling Hamiltonian as follows.
From the {\em ab-initio} data, we parameterized our Hamiltonian as:
\begin{equation}
  \hat{H}_{ab}= \sum J(u) S_i S_j + V(u),
\end{equation}
where $V(u)$ is the potential energy in the AFM state (Fig~\ref{fig:energy_as_u}).
This Hamiltonian written in second-quantized notation is
\begin{align}
  &\sum_k J(u) E_k c_k^\dagger c_k + \sum \hbar \omega_q a_q^\dagger a_q = \\
  & \sum_k J(0) E_k c_k^\dagger c_k + \sum \hbar \omega_q a_q^\dagger a_q  + \sum_k (J(u)-J(0)) E_k c_k^\dagger c_k 
\end{align}
where $E_k=Z S |\sin ka | $ is the energy of the Heisenberg model as a function of $k$\cite{coldea_spin_2001}, $\hbar \omega$ is the phonon energy, $c_k$ destroys a magnetic state, and $a_q$ destroys a phonon state.
The matrix elements of the Hamiltonian are thus
\begin{align}
  \bra{c_{k'} n_{q'}'}\hat{H}\ket{c_{k}n_{q}} 
  =& \delta_{k'k}\delta_{n_{q'}'n_q}(J(0)E_k+\hbar\omega_q n_q) \\
   & + E_k\delta_{k'k'} \bra{n_{q'}'}(J(u)-J(0))\ket{n_q},
\end{align}
where $n_q$ is the excitation level of the phonon of wave vector $q$.
Here we have just considered the $\Gamma$-point \aog and \bog modes, so we will consider the coupling for $q'=q=0$.
Since $J(u)$ is even for these phonons, it couples only with a double phonon excitation.
We define the dimensionless coupling constant $\gamma$ as
\begin{equation}
  \gamma= \frac{\bra{0}J(u)\ket{2}}{2\omega}=\frac{1}{2\sqrt{2}m\omega^2}\left(\delta c_1+\frac{3\delta c_2}{m\omega}\right).
  \label{eqn:gamma}
\end{equation}
(note that we divide by twice the freqency because $J(u)$ is even and thus couples with a double phonon excitation).
The calculated values for $\gamma$ are reported in Table~\ref{table:gamma}.
We report only the La$_2$CuO$_4$ and CaCuO$_2$ coupling constants, since they have stable phonon modes.
Both of these materials have rather large dimensionless coupling constants.


To assess the origin of the strong coupling calculated {\em ab initio}, we compare to a simple model based on electron hopping.
Following Hafliger~\cite{hafliger_quantum_2013}, the hopping between oxygen and copper is given by 
\begin{equation}
  t_{pd}=A_0 d^{-\alpha_0} \cos^{\beta_0}\theta,
\end{equation}
where $\alpha_0=3$ or $3.5$, $\beta_0=1$, $d$ is the Cu-O bond length, and $\theta$ is the Cu-O-Cu bond angle.
For direct comparison, we will write $d$ and $\theta$ in terms of the symmetric bond length $d_0\simeq 1.6\AA$ and the displacement of the oxygen atoms out of the plane $u$.
$d^2=d_0^2+u^2$, and for small distortions, one can approximate $\cos\theta\simeq -1 +2 u^2/d_0^2$.
Since $J\propto t_{pd}^4$, we obtain 
\begin{equation}
  J(u)=J_0d_0^{4\alpha_0}(d_0^2+u^2)^{-2\alpha_0}\left(1-\frac{2u^2}{d_0^2}\right)^4
  \label{eqn:theory_ju}
\end{equation}
where we fixed $J(0)=J_0$.
Therefore, the curvature at $u=0$ is given by $-4J_0(4+\alpha_0)d_0^{-2}$.

The hopping model allows us to assess whether there are additional physics beyond the different frequencies of the phonon modes and a traditional superexchange/hopping understanding of magnetism.
As one can see in Table~\ref{table:gamma}, the hopping model appears to be qualitatively correct, but can have errors as large as a factor of two from the fully {\em ab-initio} coupling constants.

\begin{table}
  \begin{tabular}{l|c|c}
             & Calculated $\gamma$ & Theory $\gamma$ \\
    \hline
    \hline
    LCO \bog & -.12(1) & -.14 \\
    LCO \aog & -.03(1) & -.07 \\
    CCO \bog & -1.8(1) & -2.33 \\
    CCO \aog & -.04(1) & -.17 \\
  \end{tabular}
  \caption{The coupling constant $\gamma$ from Eqn~\ref{eqn:gamma} for the two realistic cuprates considered in this work. The 'theory' numbers are calculated using $J(u)$ from Eqn~\ref{eqn:theory_ju} and the frequencies from FN-DMC. }
  \label{table:gamma}
\end{table}

\begin{figure*}
  \begin{tabular}{lccc}
    & Tetragonal & A$_{1g}$ & B$_{1g}$ \\
    La$_2$CuO$_4$ & 
    \includegraphics[width=0.5\columnwidth]{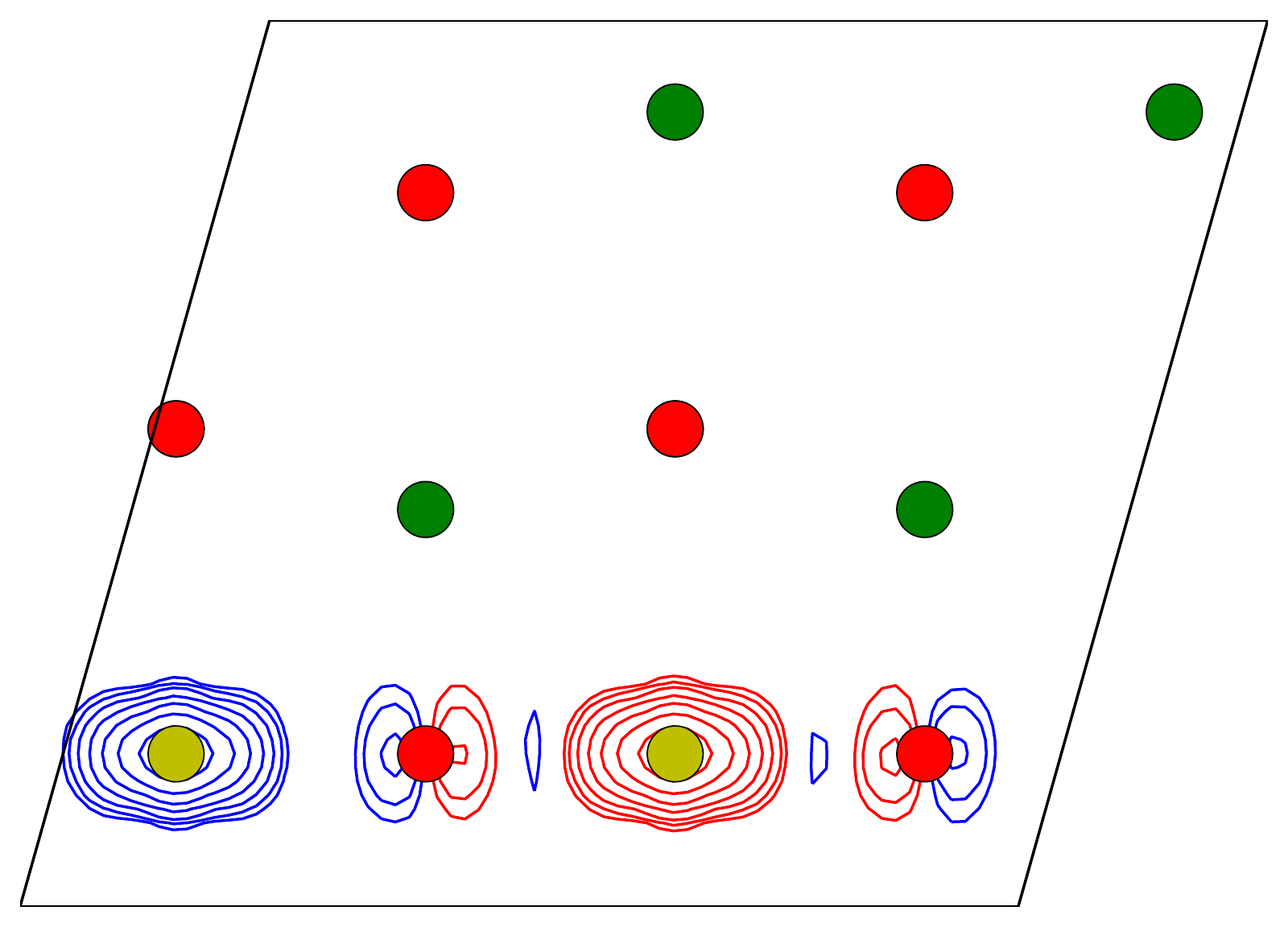} & 
    \includegraphics[width=0.5\columnwidth]{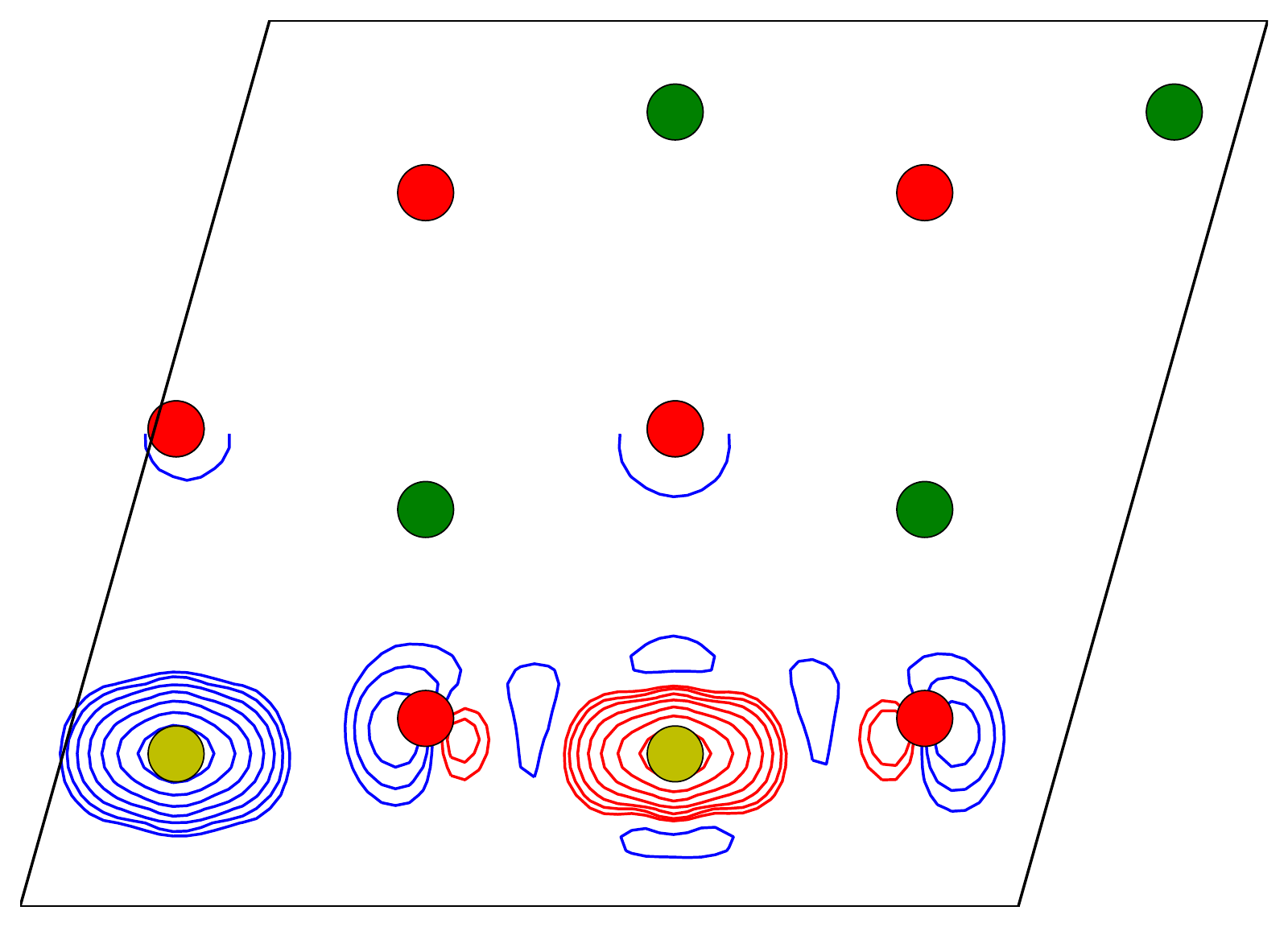} & 
    \includegraphics[width=0.5\columnwidth]{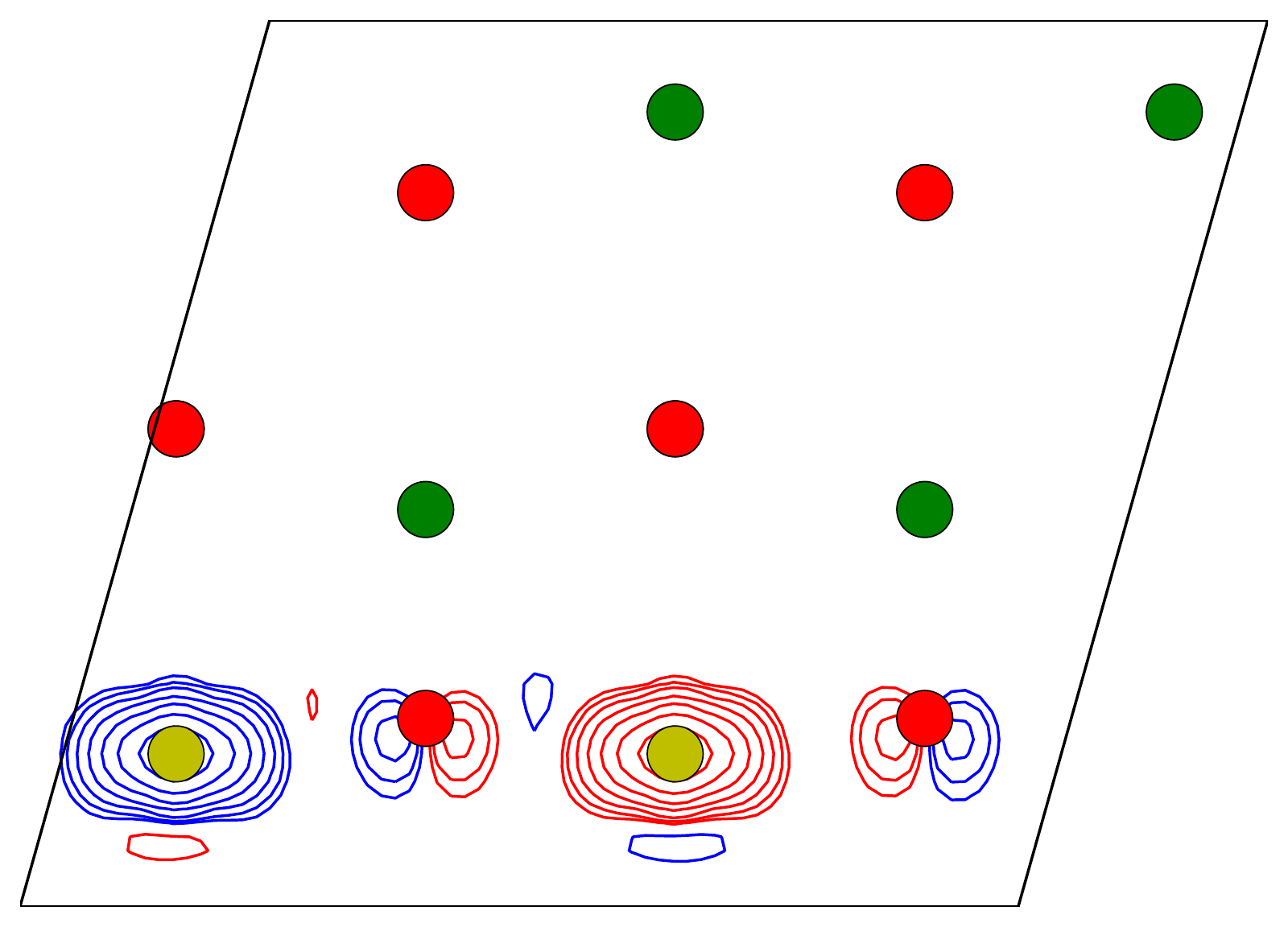} \\ 
    CaCuO$_2$ & 
    \includegraphics[width=0.5\columnwidth]{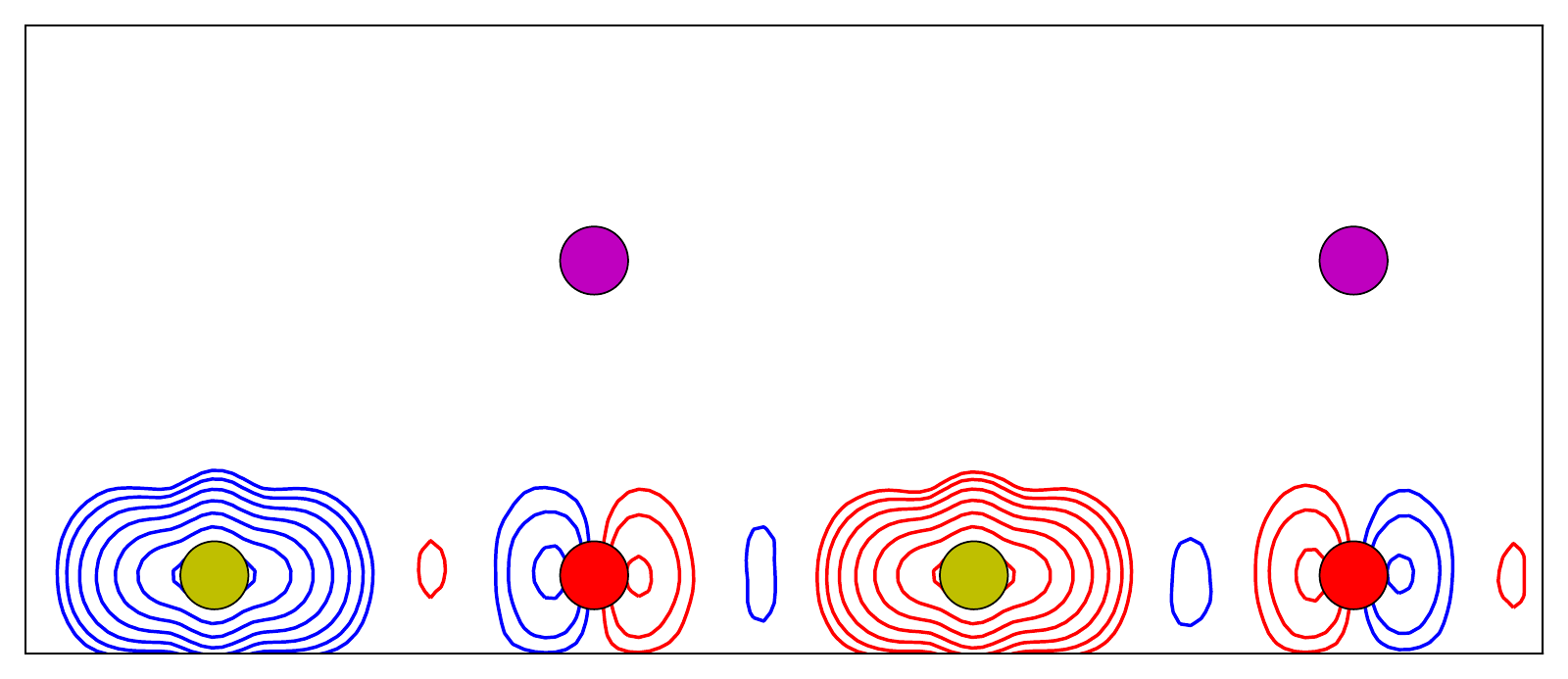} & 
    \includegraphics[width=0.5\columnwidth]{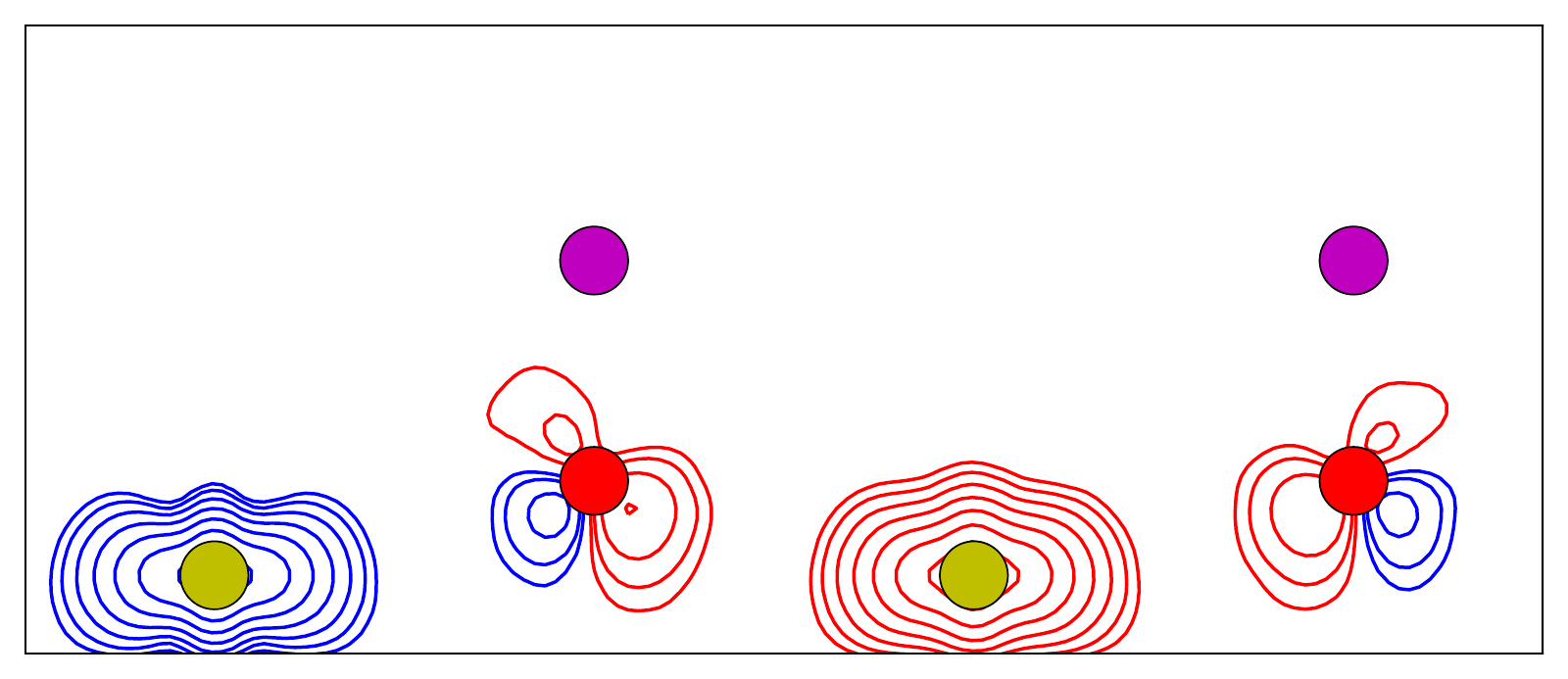} & 
    \includegraphics[width=0.5\columnwidth]{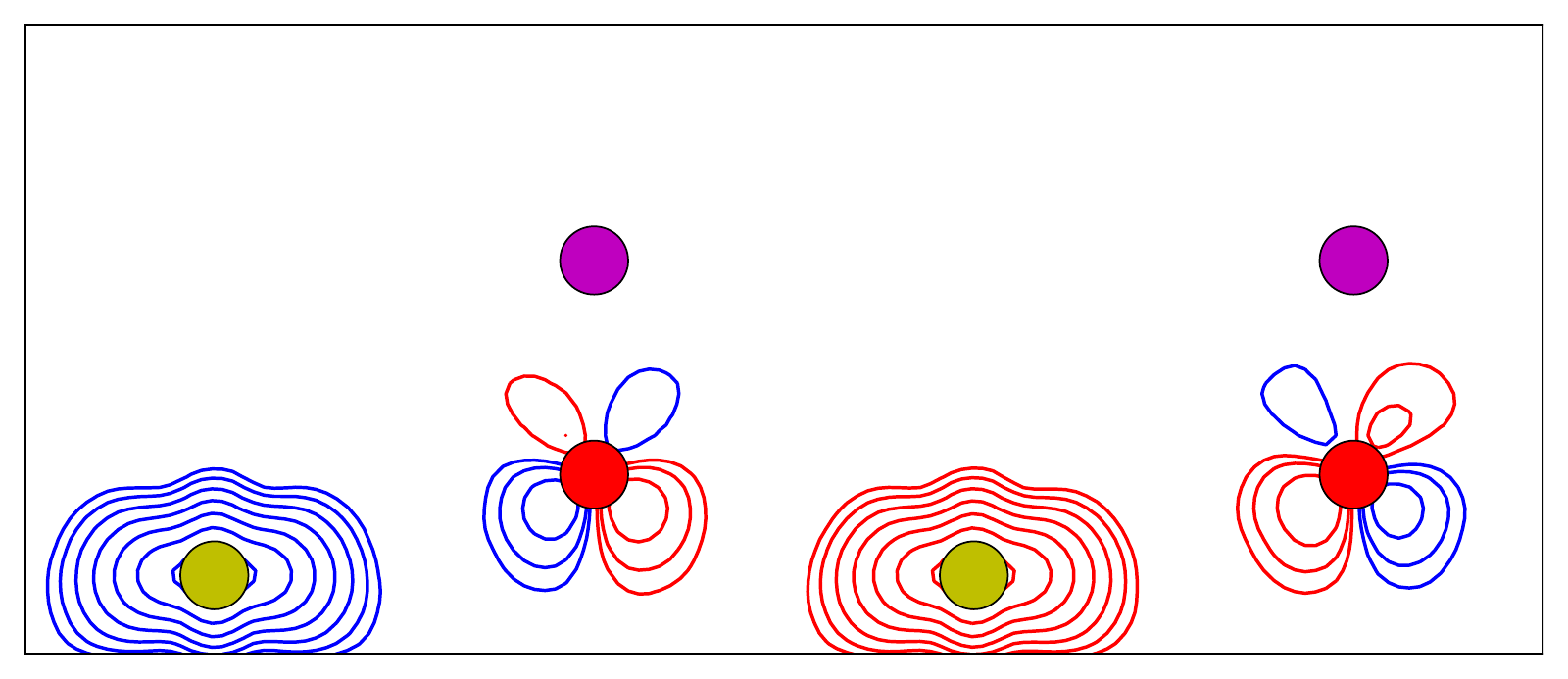}  \\
    CuO$_2^{2-}$ & 
    \includegraphics[width=0.5\columnwidth]{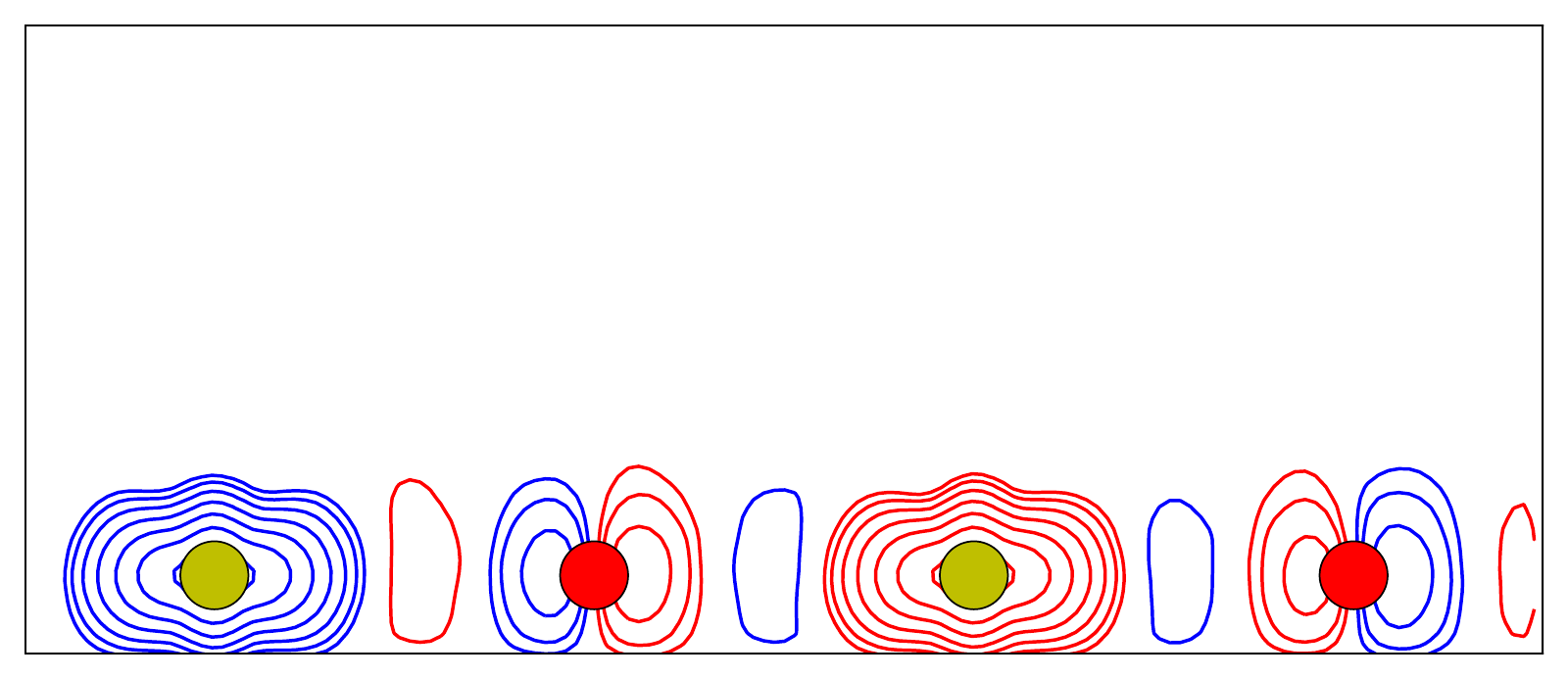} & 
    \includegraphics[width=0.5\columnwidth]{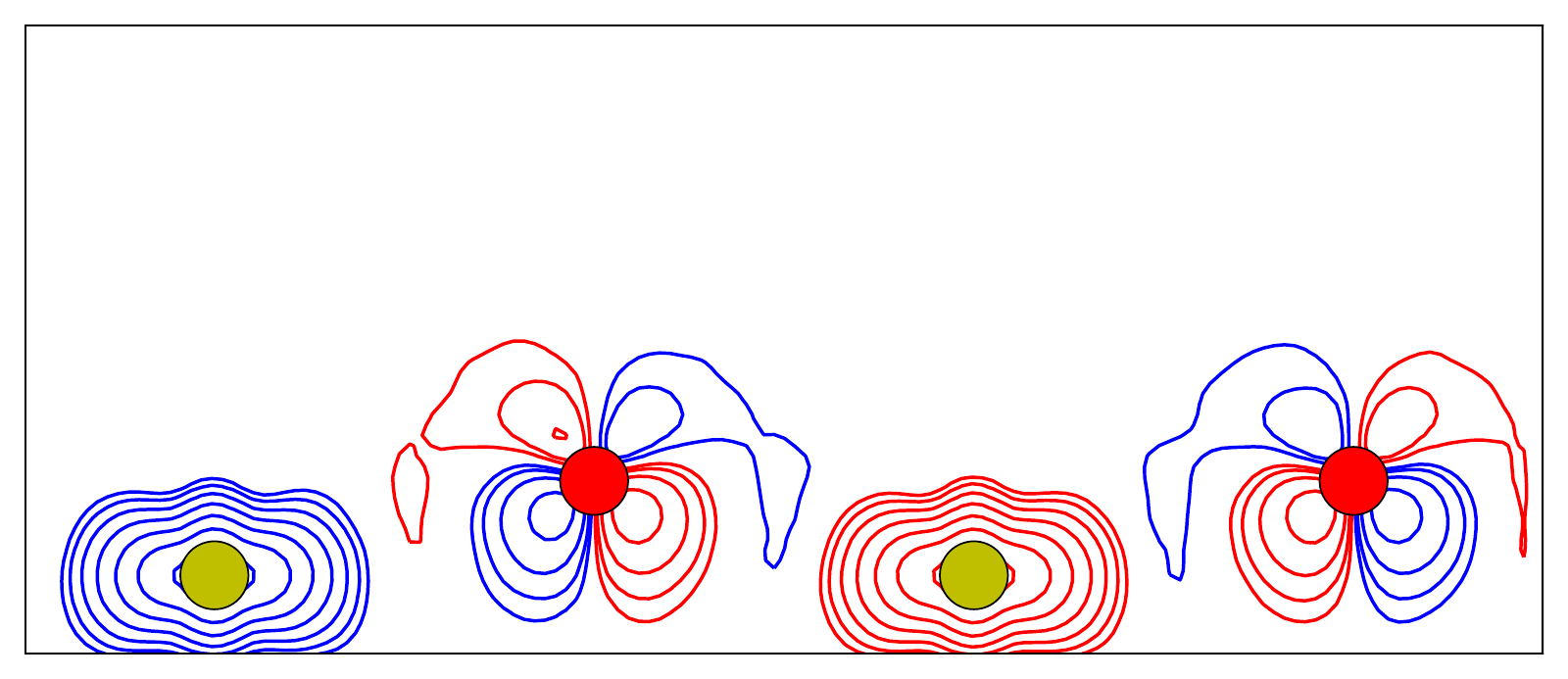} & 
    \includegraphics[width=0.5\columnwidth]{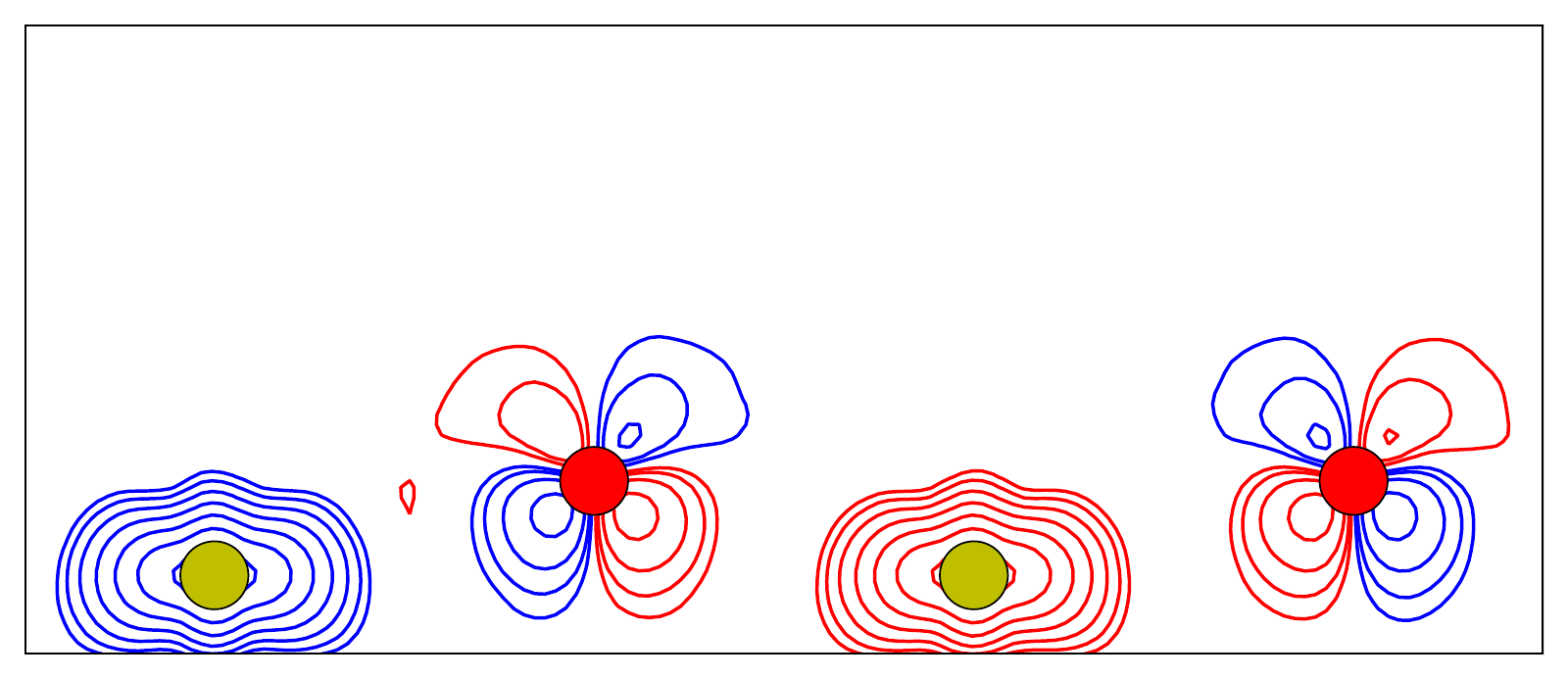}  \\
  \end{tabular}
  \caption{A slice through the $ac$ plane of the three cuprates considered in this study.  The column marked `Tetragonal' denotes the undistorted P4/mmm structure, while the \aog and \bog columns denote frozen phonon vibrations with each of those symmetries. Contour lines are logarithmically spaced, with red denoting down spin and blue up spin. The projected position of the atoms are denoted by circles: Cu(yellow), O(red), Ca(magenta), and La(green).}
  \label{fig:density} 
\end{figure*}

We thus have a clear breakdown of the strong \bog magneto-structural coupling.
The \aog mode has a higher frequency for two main reasons: the first is that it interacts strongly with the interlayer, and second, that the bond angle of the copper for the \aog mode is more unfavorable; the \bog mode creates a tetrahedral arrangement, which is energetically more accessible.
This can be seen in the curvatures for different cuprates in Fig~\ref{fig:energy_as_u}, and in the spin density plots in Fig~\ref{fig:density}, in which it is clear that the copper-oxygen bonding and superexchange are both affected by the interlayer, once the phonon mode is activated.
Secondly, a simple hopping model for the superexchange tends to {\em overestimate} the coupling given the difference in frequency, presumably because the interlayer disrupts the superexchange, as we see in Fig~\ref{fig:density}.
The main effect appears to be the difference in frequencies, which results in differences of an order of magnitude in the coupling, with the beyond-hopping physics occuring at the level of a factor of two.

The picture presented here fits well with experimental data.
In LCO, the frequency of the \bog mode is changed by 6-8 meV upon changing the magnetic state, which is 18-24\% of the frequency of the mode.
This is of a size large enough to explain the shift in the \bog mode observed in experiment\cite{reznik_q_1995}.
By calculating the expectation value of $J(u)$ on the phonon wave function for different atomic masses, 
we find that the calculated isotope effect on $J$ is negligibly small, which is also observed in experiment\cite{zhao_oxygen_1994}.
Finally, the magneto-structural coupling allows for the existence of phonon side bands in the magnetic spectrum, which have been observed\cite{perkins_mid-infrared_1993,lorenzana_phonon_1995}.

In summary, the results presented here are two-fold.  
The first is that we have demonstrated that the base state of the cuprates can be described accurately with a fully first-principles implementation of quantum Monte Carlo techniques.
Since this method has no adjustable parameters, it is predictive and can be used in searches for new exotic materials.
In addition, since we calculate rather than presuppose the electronic correlations, the method can be used to study electron correlation on an even footing with one-body effects.
The second main result is that the coupling between magnetism and the lattice are quite large.  For the \bog mode in LCO, this is close to what one would expect from a simple hopping theory.
However, in the \aog mode, the interlayer prevents the magneto-structural coupling from occuring, mainly by shifting the phonon frequency up, but also partially by disrupting the antiferromagnetic exchange pathways.  This mechanism may explain why experiment observes a shift in the \bog mode but not the \aog mode upon entering the superconducting state.

The results contained herein emphasize the importance of treating the electron correlations explicitly and on an equal footing to the one-body effects in a simulation of strongly interacting systems like the cuprates.
Even one-body properties such as the delocalization are affected by electron correlation, and cannot be taken at face value from a density functional theory calculation.
The FN-DMC method, with modern techniques, is so far able to cleanly connect the first-principles Hamiltonian to observed phenomena in these materials, without artificially adding terms to account for their strongly correlated nature.
This new capability in electronic structure calculations has tremendous potential to provide a detailed microscopic description of the physics of these challenging many-body systems.

We would like to acknowledge many useful conversations with David Ceperley, Laura Greene, Jim Eckstein, Jeremy Morales, and Brian Busemeyer.
The authors gratefully acknowledge funding from DOE FG02-12ER46875
and computer resources from the Blue Waters friendly user period, PRAC JMP award,  and Illinois JPL award.
P.A. was supported by DOE grant DE-FG02-06ER46285.

\bibliography{cuprate_paper}

\end{document}